\documentclass[nonacm,sigplan,screen]{acmart}

\usepackage{fontawesome}
\usepackage{hyperref}
\usepackage[capitalise,noabbrev]{cleveref}
\usepackage{url}
\usepackage{caption}
\usepackage{subcaption}
\usepackage{graphicx}
\usepackage{multirow}
\usepackage{listings}
\usepackage{framed}
\usepackage{amsmath}
\usepackage{color}
\usepackage{xcolor}
\usepackage{tikz}
\usepackage{enumitem}
\usepackage{threeparttable}

\lstdefinelanguage[RISC-V]{Assembler}
{
  alsoletter={.}, 
  alsodigit={0x}, 
  morekeywords=[1]{ 
    lb, lh, lw, ld, lbu, lhu, lwu,
    sb, sh, sw, sd,
    sll, slli, srl, srli, sra, srai,
    add, addi, sub, lui, auipc,
    xor, xori, or, ori, and, andi,
    slt, slti, sltu, sltiu,
    beq, bne, blt, bge, bltu, bgeu,
    j, jr, jal, jalr, ret,
    ecall, ebreak,
    nop, la, li
  },
  morekeywords=[2]{ 
    .align, .ascii, .asciiz, .byte, .data, .double, .extern,
    .float, .globl, .half, .kdata, .ktext, .set, .space, .text, .word
  },
  morekeywords=[3]{ 
    zero, ra, sp, gp, tp, s0, fp,
    t0, t1, t2, t3, t4, t5, t6,
    s1, s2, s3, s4, s5, s6, s7, s8, s9, s10, s11,
    a0, a1, a2, a3, a4, a5, a6, a7,
    ft0, ft1, ft2, ft3, ft4, ft5, ft6, ft7,
    fs0, fs1, fs2, fs3, fs4, fs5, fs6, fs7, fs8, fs9, fs10, fs11,
    fa0, fa1, fa2, fa3, fa4, fa5, fa6, fa7
  },
  morecomment=[l]{;},   
  morecomment=[l]{\#},  
  morestring=[b]",      
  morestring=[b]'       
}

\definecolor{mauve}{rgb}{0.58,0,0.82}
\definecolor{taint}{RGB}{197, 41, 119}

\newcommand{\policysym}[1]{\mkern1.5mu{#1}\mkern1.5mu}
\newcommand{\trippleop}[3]{#1\policysym{?}#2\policysym{:}#3}
\newcommand{\diffsig}[1]{\textcolor{taint}{#1_{\mathit{diff}}}}
\newcommand{\rvcomment}[1]{\textcolor{green!50!black}{$#1$}}

\definecolor{vgreen}{RGB}{104,180,104}
\definecolor{vblue}{RGB}{49,49,255}
\definecolor{vorange}{RGB}{255,143,102}
\lstdefinestyle{verilog-style}
{
    language=Verilog,
    basicstyle=\small\ttfamily,
    keywordstyle=\color{vblue},
    identifierstyle=\color{black},
    commentstyle=\color{vgreen},
    numbers=left,
    numberstyle=\tiny\color{black}
}

\crefformat{section}{\S#2#1#3}
\crefformat{equation}{Policy #2#1#3}
\newcommand{\keypoint}[1]{\noindent\textbf{#1}}
\newcommand{\code}[1]{\texttt{{\small \detokenize{#1}}}}

\newcommand{\name}{DejaVuzz}
\newcommand{\challenge}[1]{\textbf{C{#1}}}
\newcommand{\circled}[1]{\textcircled{\raisebox{-0.8pt}{#1}}}
\newcommand\bcircled[1]{\tikz[baseline=(char.base)]{
            \node[shape=circle,fill,inner sep=0.6pt] (char) {\textcolor{white}{#1}};}}
\newcommand{\introspec}{\textsc{IntroSpectre}}
\newcommand{\specdoc}{\textsc{SpecDoctor}}
\newcommand{\teesec}{\textsc{TEESec}}
\newcommand{\cellift}{\textsc{CellIFT}}
\newcommand{\diffift}{diffIFT}
\newcommand{\swapmem}{swapMem}

\begin{document}

\title{\name{}: Disclosing Transient Execution Bugs with Dynamic Swappable Memory and Differential Information Flow Tracking assisted Processor Fuzzing}

\author{Jinyan Xu}
\affiliation{
  \institution{Zhejiang University}
  \city{Hangzhou}
  \state{Zhejiang}
  \country{China}
}
\email{phantom@zju.edu.cn}

\author{Yangye Zhou}
\affiliation{
  \institution{Zhejiang University}
  \city{Hangzhou}
  \state{Zhejiang}
  \country{China}
}
\email{zhouyangye@zju.edu.cn}

\author{Xingzhi Zhang}
\affiliation{
  \institution{Zhejiang University}
  \city{Hangzhou}
  \state{Zhejiang}
  \country{China}
}
\email{xingzhizhang@zju.edu.cn}

\author{Yinshuai Li}
\affiliation{
  \institution{Southern University of Science and Technology}
  \city{Shenzhen}
  \state{Guangdong}
  \country{China}
}
\email{liys2022@mail.sustech.edu.cn}

\author{Qinhan Tan}
\affiliation{
  \institution{Princeton University}
  \city{Princeton}
  \state{New Jersey}
  \country{USA}
}
\email{qinhant@princeton.edu}

\author{Yinqian Zhang}
\affiliation{
  \institution{Southern University of Science and Technology}
  \city{Shenzhen}
  \state{Guangdong}
  \country{China}
}
\email{yinqianz@acm.org}

\author{Yajin Zhou}
\affiliation{
  \institution{Zhejiang University}
  \city{Hangzhou}
  \state{Zhejiang}
  \country{China}
}
\email{yajin_zhou@zju.edu.cn}

\author{Rui Chang}
\affiliation{
  \institution{Zhejiang University}
  \city{Hangzhou}
  \state{Zhejiang}
  \country{China}
}
\email{crix1021@zju.edu.cn}

\author{Wenbo Shen}
\affiliation{
  \institution{Zhejiang University}
  \city{Hangzhou}
  \state{Zhejiang}
  \country{China}
}
\email{shenwenbo@zju.edu.cn}

\begin{abstract}
Transient execution vulnerabilities have emerged as a critical threat to modern processors.
Hardware fuzzing testing techniques have recently shown promising results in discovering transient execution bugs in large-scale out-of-order processor designs.
However, their poor microarchitectural controllability and observability prevent them from effectively and efficiently detecting transient execution vulnerabilities.

This paper proposes \name{}, a novel pre-silicon stage processor transient execution bug fuzzer.
\name{} utilizes two innovative operating primitives: dynamic swappable memory and differential information flow tracking, enabling more effective and efficient transient execution vulnerability detection.
The dynamic swappable memory enables the isolation of different instruction streams within the same address space.
Leveraging this capability, \name{} generates targeted training for arbitrary transient windows and eliminates ineffective training, enabling efficient triggering of diverse transient windows.
The differential information flow tracking aids in observing the propagation of sensitive data across the microarchitecture.
Based on taints, \name{} designs the taint coverage matrix to guide mutation and uses taint liveness annotations to identify exploitable leakages.
Our evaluation shows that \name{} outperforms the state-of-the-art fuzzer \specdoc{}, triggering more comprehensive transient windows with lower training overhead and achieving a 4.7$\times$ coverage improvement.
And \name{} also mitigates control flow over-tainting with acceptable overhead and identifies 5 previously undiscovered transient execution vulnerabilities (with 6 CVEs assigned) on BOOM and XiangShan.
\end{abstract}

\maketitle 
\pagestyle{plain} 

\section{Introduction}

The recent discovery of transient execution vulnerabilities has unveiled a significant threat to modern processors.
These vulnerabilities, such as Spectre~\cite{kocher2020spectre} and Meltdown~\cite{lipp2020meltdown}, exploit speculative execution, a key performance optimization feature, to leak sensitive data through side channels.
The ongoing battle between attackers and defenders resembles a continuous cat-and-mouse game.
For example, Spectre-V2~\cite{kocher2020spectre} promoted the privilege-isolated branch prediction deployment, but follow-up research soon discovered bugs~\cite{enrico2022bhi, wikner2023phantom, trujillo2023inception} in other speculation components. 
Similarly, after Foreshadow~\cite{vanbulck2018foreshadow} was patched, Microarchitectural Data Sampling (MDS)~\cite{ridl, canella2019fallout} attacks emerged.
This arms race not only challenges the efficacy of existing defense mechanisms but also underscores the necessity of a proactive approach to automated transient execution bug detection.

Some efforts~\cite{yuan2020speechminer, moghimi2020medusa, oleksenko2022revizor} have been applied to commodity processors.
However, due to the black-box nature of off-the-shelf processors, these approaches rely heavily on template-based generation and fixed side channels, which makes it difficult for them to uncover new vulnerabilities.
On the contrary, detection approaches at the pre-silicon stage have yet to be extensively studied.
Detecting these vulnerabilities during the Register Transfer Level (RTL) development phase is crucial, as hardware bugs are usually difficult to fix once the design is manufactured.
Early detection allows for timely remediation, preventing these bugs from being integrated into production hardware.
Therefore, proactive testing and verification at the pre-silicon stage is imperative for ensuring processor microarchitecture security.

Formal verification and fuzzing are commonly used methods for existing processor RTL transient execution bug detection.
Although formal approaches can prove security properties exhaustively, limited by the state explosion problem, existing methods~\cite{yang2023pensieve,trippel2018checkmate,tan2024rtl,fadiheh2022exhaustive} solve the scalability problem by modeling processor transient execution behavior at a higher level of abstraction.
However, given the complexity of the out-of-order processor design, the microarchitecture implementation details ignored by the model are highly error-prone~\cite{morfuzz, specdoctor}.
Furthermore, the complicated design pre-knowledge and heavy manual efforts required for hardware modeling and security property definition also impede applying formal methods to complex designs.

Recently, processor fuzzing has demonstrated promising results in verifying large-scale complex processor designs~\cite{difuzzrtl, thehuzz, morfuzz, hypfuzz, cascade}, and researchers also have begun applying fuzzing to detect transient execution vulnerabilities~\cite{introspectre, specdoctor, teesec}.
\introspec{}~\cite{introspectre} and \teesec{}~\cite{teesec} use gadget templates to generate Meltdown-type transient execution vulnerabilities and identify leakage by searching for secret values in the microarchitecture logs.
\specdoc{}~\cite{specdoctor}, on the other hand, employs a multi-phase random instruction generation process and utilizes differential testing to detect sensitive data leakage.
However, due to the complexity of the transient execution vulnerabilities, current fuzzing methods are either too limited~\cite{introspectre, teesec}, only capable of identifying specific leakage patterns, or too inefficient~\cite{specdoctor}, taking days to complete the detection, thereby limiting their practical adoption.
To effectively and efficiently fuzz transient execution bugs, the following two challenges need to be addressed.

First, only transiently executed instructions are considered effective fuzzing payloads, so the fuzzer needs to efficiently trigger diverse transient windows for fuzzing.
However, triggering these transient windows requires deliberate microarchitecture training.
Due to significant differences in training patterns among various microarchitecture components, existing approaches generate limited transient windows with high training overhead (\cref{eval:control}).
The inability to generate various transient windows means the microarchitecture cannot be fully explored.
Additionally, ineffective training instructions waste simulation time, increasing training overhead and reducing the fuzzing throughput.

Second, the fuzzer needs to perceive the propagation of sensitive data during transient execution to guide mutation and detect leakages.
Information flow tracking is a promising solution, but it suffers from the control flow over-tainting problem in complex designs~\cite{cellift}.
Due to the lack of effective methods to trace sensitive data, existing fuzzers cannot measure coverage or identify exploitable leakages (\cref{eval:observe}).
Lacking coverage metrics means that the quality of stimuli cannot be assessed, leading to inefficient input mutation. 
Passing unexploitable leakages to subsequent stages not only results in false positives but also makes later phases futile, further misguiding the fuzzing process.

To address the challenges mentioned, we propose \name{}, an effective and efficient pre-silicon processor fuzzer for transient execution vulnerabilities, powered by two novel operating primitives: dynamic swappable memory and differential information flow tracking.
Dynamic swappable memory serves as an isolation primitive, responsible for transparently switching instruction sequences to control the microarchitecture to trigger desired transient execution behaviors.
This primitive resolves conflicts between instruction sequences by time-sharing the address space.
To increase the diversity of triggered transient windows, \name{} isolates training and transient instruction sequences to generate arbitrary transient windows, and uses the training derivation strategy to derive targeted training based on transient execution information.
To reduce the training overhead, \name{} isolates each training instruction sequence to explore different training effects, and eliminates ineffective training through the training reduction strategy.
%
Differential information flow tracking acts as the tracing primitive that is responsible for observing microarchitecture state changes caused by sensitive data.
This primitive eliminates the control flow over-tainting problem by comparing whether different secrets can produce different selections on the same control signal.
With the help of taints, \name{} designs a taint coverage matrix to evaluate how sensitive data propagates during the transient execution, effectively guiding exploration.
Furthermore, \name{} introduces taint liveness annotations to bind state registers to related taint registers.
By using annotated state registers as liveness signals, \name{} filters out unexploitable taints to reduce false positives.

Overall, this paper makes the following contributions:

\begin{itemize}[noitemsep, leftmargin=*]
\item We summarize the challenges of transient execution bug fuzzing in terms of microarchitectural controllability and microarchitectural observability and propose two novel operating primitives:
a \textit{dynamic swappable memory} model to resolve address space conflicts for better microarchitectural control,
and a \textit{differential information flow tracking} technique to mitigate control flow over-tainting for improved microarchitectural observation.
\item Utilizing these two operating primitives, we develop a new processor fuzzing framework named \name{}, which effectively and efficiently detects transient execution bugs.
\name{} designs training derivation and training reduction strategies atop dynamically swappable memory to efficiently trigger diverse transient windows, and utilizes taints generated by differential information flow tracking to guide fuzzing and identify leakage.
\item We evaluate \name{} on two well-known RISC-V out-of-order processors~\cite{boom, xiangshan}.
Compared to the SOTA fuzzer \specdoc{}~\cite{specdoctor}, \name{} achieves a 4.7$\times$ improvement in coverage with more comprehensive transient windows and lower training overhead.
\name{} mitigates control flow over-tainting with acceptable overhead and identifies 5 previously unknown transient execution vulnerabilities, all of which are assigned CVE numbers.

\end{itemize}

To facilitate the community and future research, we publish the source code and experiments of \name{} at \url{https://github.com/sycuricon/DejaVuzz}.


\section{Background}


\subsection{Transient Execution Vulnerabilities}
\label{sec:bug}

\begin{figure}[t]
    \centering
    \includegraphics[width=\linewidth]{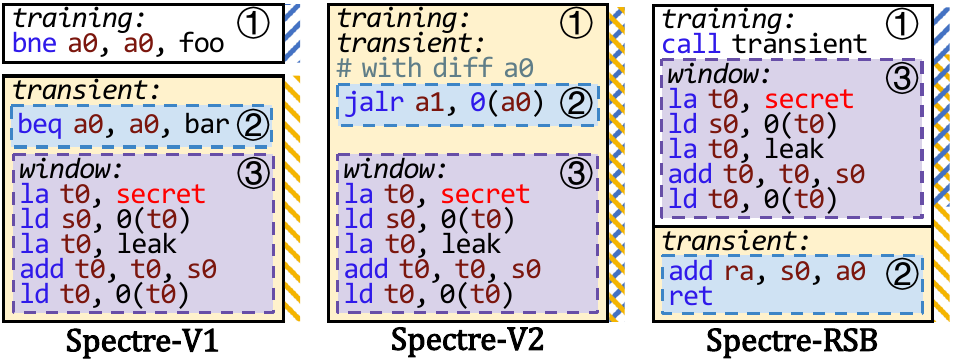}
    \caption{Training and transient execution sections of Spectre-V1, Spectre-V2 and Spectre-RSB. The secret decoding step \circled{4} is omitted.}
    \label{fig:bug}
\end{figure}

As shown in \cref{fig:bug}, the process of exploiting a transient execution bug can be divided into the following 4 attack steps:
\circled{1} training the target microarchitecture, \circled{2} triggering a transient window through the trained state, \circled{3} accessing sensitive data and encoding it into a side channel, and \circled{4} subsequently decoding the secret from the side channel.

However, different types of transient windows exhibit highly varied training patterns.
For Spectre-V1, the training section (blue stripe) and the transient execution section (yellow stripe) are independent, which means these two sections can be generated independently as long as the branch instructions have the same address offset.
However, this is not always true for other transient execution bugs such as Spectre-V2 and Spectre-RSB~\cite{koruyeh2018spectre, maisuradze2018ret2spec}.
The Spectre-V2 attack requires different arguments (\code{a0}) to switch between training and exploiting the Branch Target Buffer (BTB) with the same code.
And the Spectre-RSB attack requires tempting the processor to speculatively return to a corrupt address by training the Return Stack Buffer (RSB).
As seen in the last two types, complex transient windows are mixed with the training section.
Triggering such complex transient windows is challenging, as the stimulus generator must carefully handle the semantics of training and transient execution to ensure the \code{window} section is executed transiently as expected.
Otherwise, non-speculative execution of the \code{window} section during training could lead to false positives.


\subsection{Hardware Dynamic Information Flow Tracking}
\label{sec:ift}

\begin{figure}[t]
    \centering
    \includegraphics[width=0.9\linewidth]{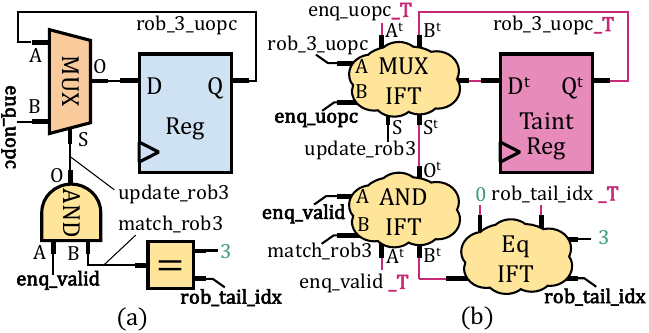}
    \caption{Hardware dynamic information flow tracking instrumentation. a) is the example circuit from the BOOM RoB module, b) is the corresponding IFT shadow circuit.}
    \label{fig:ift}
\end{figure}

Information Flow Tracking (IFT) has been widely deployed at all levels of hardware abstraction to understand how information flows through a system~\cite{tiwari2011crafting, li2011caisson, weisse2019nda, iftsurvey}.
Hardware dynamic IFT, known as taint tracking, can dynamically verify information flow properties during the runtime.
This is achieved by marking sensitive state elements with taints at the circuit level and propagating the taints based on the operations on sensitive data.
There are three instrumentation levels for the hardware dynamic IFT mechanism: gate level~\cite{glift}, RTL level~\cite{rtlift}, and cell level~\cite{cellift}. \cref{fig:ift} shows how hardware dynamic IFT is implemented in hardware.
The dynamic IFT instrumentation generates a shadow circuit based on the original circuit, all registers in the original circuit are copied to store taints, and the combinational logic gates are replaced with the corresponding taint propagation policy implementation.
The taint propagation policies are a set of rules that are responsible for tainting outputs that are affected by tainted inputs.
Policies \ref{rule:and} and \ref{rule:mux} are the state-of-the-art taint propagation policies~\cite{rtlift, cellift} for the AND and MUX cells, respectively.
By using shadow circuits, dynamic IFT provides the ability to observe the information flow of the design without affecting the original functionality.


\vspace{-8pt}
\begin{equation}
    O^{t}_{AND} = (A\policysym{\&}B^{t}) | (B\policysym{\&}A^{t}) | (A^{t}\policysym{\&}B^{t})
    \label{rule:and}
\end{equation}

\vspace{-10pt}
\begin{equation}
    O^{t}_{MUX} = (\trippleop{S}{B^{t}}{A^{t}}) | (\underline{\trippleop{S^{t}}{(A \textasciicircum B)|(A^{t} | B^{t})}{0}})
    \label{rule:mux}
\end{equation}

Taints generated by the direct computation of input taints and signals, like in \cref{rule:and}, are referred to as data taints.
In \cref{rule:mux}, in addition to selecting data taints via the selection signal \code{S}, the underlined component produces control taints due to the conditional selection semantics of the multiplexer.
Unlike data taints, which are only impacted by the actually executed code, control taints also consider changes occurring on unselected branches (i.e., the $A \textasciicircum B$ term).
Thus, once taints propagate to the control flow, it can easily lead to over-tainting~\cite{schwartz2010all, cellift}.
Since taint propagation policies only generate taints without eliminating them, more registers become tainted as the circuit executes, making it increasingly difficult to identify target information flows precisely.

According to our evaluation (\cref{eval:observe}), the state-of-the-art hardware dynamic IFT mechanism \cellift{}~\cite{cellift} suffers from the control flow over-tainting problem.
Next, we use the Reorder Buffer (RoB) module of BOOM~\cite{boom} in \cref{fig:ift} as an example to explain how the taint explosion occurs during the RoB rollback.
The third RoB entry updates its opcode field register \code{rob_3_uopc} with the new opcode \code{enq_uopc} when a valid micro-operation is enqueued (\code{enq_valid} is high) and the tail pointer points to the third entry (\code{rob_tail_idx} is equal to 3).
Before the RoB rollback, instructions using tainted sensitive data as operands in step \circled{3} write back and taint the RoB state register. 
When the RoB rolls back, the movement of the tail pointer causes \code{rob_tail_idx} to be tainted.
Since the frontend also uses the RoB index to maintain state, \code{enq_valid} is tainted.
According to \cref{rule:and}, both inputs are tainted (the comparison result of the \code{Equal} cell is also tainted due to the tainted \code{rob_tail_idx}), causing the MUX selection signals to be marked as tainted.
Furthermore, based on \cref{rule:mux}, the register \code{rob_3_uopc} is also marked as tainted due to the different input data.
All 736 RoB entry field registers have a similar update logic. Therefore, they are all suddenly tainted when the RoB rolls back.

\subsection{Processor Fuzzing for Transient Execution Bugs}

Processor fuzzing has been employed to detect various bugs, including functional bugs~\cite{difuzzrtl, thehuzz, morfuzz, cascade}, transient execution bugs~\cite{introspectre, teesec, specdoctor}, and side-channel bugs~\cite{sigfuzz}.
Although bugs are characterized differently, existing fuzzers generally follow a similar workflow consisting of three main steps.

First, the input generator generates instruction sequences as stimuli either based on constraints~\cite{morfuzz, cascade} or through random generation~\cite{difuzzrtl, thehuzz, sigfuzz}.
As discussed in \cref{sec:bug}, a transient execution attack involves multiple steps.
Thus, existing fuzzers strategically divide the generation into multiple phases.
For instance, \introspec{} and \teesec{} complement the main gadget with the preceding helper gadgets depending on whether the target memory access paths are met in the software execution model.
\specdoc{} sequentially progresses through the transient-trigger, secret-transmit, and secret-receive phases to generate a complete stimulus.
During each phase, additional instructions are randomly appended to those generated in the previous phase until specific goals are met.
The goals of each phase are to trigger a RoB rollback, generate differences in microarchitecture, and cause differences in execution cycles.

Second, the fuzzer uses an RTL simulator to convert the Design Under Test (DUT) into a software model and then uses the model to execute the generated instruction sequences.
During simulation, the fuzzer leverages instrumentation to measure coverage to guide mutations.
Existing fuzzers define several coverage metrics to reflect the general processor behavior, such as mux toggle coverage~\cite{rfuzz}, control register coverage~\cite{difuzzrtl, morfuzz}, or hardware behavior coverage~\cite{thehuzz}.
However, transient execution vulnerabilities focus more on propagating sensitive data within the microarchitecture.
Therefore, existing general processor behavior coverage metrics are unsuitable for transient execution vulnerabilities.

Third, the fuzzer analyzes the microarchitecture to determine if any bug exists.
Unlike the functional bugs that can be detected using co-simulation~\cite{difuzzrtl, morfuzz}, transient execution vulnerabilities require detailed microarchitecture analysis.
For example, \introspec{} and \teesec{} dump the microarchitecture at each cycle and then assess whether leakage has occurred based on the presence of the secret values in the log.
\specdoc{} observes execution behavior by hashing the final state of the timing components after transient execution and evaluates leakage by comparing the consistency of the hash values between different variants.

\section{Operating Primitives}


In this section, we first analyze the challenges of transient execution fuzzing based on the key capabilities required by a fuzzer and identify their root causes.
Next, we present the design of two novel operating primitives and explain how they address the root causes.
For the challenges, we use designs based on the primitives to address them in \cref{sec:design}.

\subsection{Challenges and Root Causes}
\label{sec:challenge}


The task of a transient execution bug fuzzer is to generate instruction sequences that trigger transient windows and encode secrets into the microarchitecture, and then determine whether the encoded states can leak the secrets.
To achieve this, a competent fuzzer must possess two key capabilities.
First, it must effectively train the microarchitecture to trigger diverse transient windows, since we are only interested in transiently executed behaviors.
Second, it must accurately track the propagation of sensitive data, as we only focus on microarchitecture changes caused by secrets.
Based on this observation, we define these two capabilities as microarchitectural controllability and observability, respectively.

\begin{figure}[t!]
    \centering
    \includegraphics[width=\linewidth]{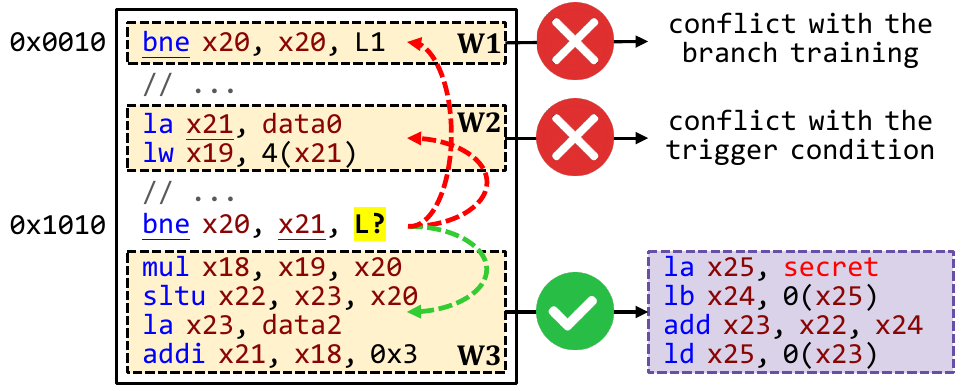}
    \caption{
    Assuming the branch instruction at \code{0x1010} can trigger transient windows at different addresses by using different branch targets \code{L?}, only transient windows that do not conflict with training instructions can be exploited.
    }
    \label{fig:conflict}
\end{figure}

\textbf{Microarchitectural Controllability}~\cite{oleksenko2020specfuzz, wiebing2024inspectre, easdon2022rapid} refers to the ability of a fuzzer to efficiently manipulate microarchitecture to trigger desired transient execution behaviors.
Existing fuzzers generate transient windows using template-based~\cite{introspectre, teesec} or random-based~\cite{specdoctor} methods.
While they can successfully trigger transient windows, they fail to address the following two challenges.

\keypoint{\challenge{1-1}. Limited Transient Window.}
Template-based methods are limited to specific transient window templates, while random-based methods also fail to generate arbitrary transient windows.
As shown by W3 in \cref{fig:conflict}, \specdoc{} randomly generates training instructions and replaces the RoB squashed instructions with the secret encoding instructions to exploit.
However, when the RoB squashed instructions are mixed with the training instructions (i.e., complex transient windows in \cref{sec:bug}), replacing them may invalidate transient execution.
For example, replacing branch training can prevent the predictor from reaching the desired prediction state (W1), while replacing the assignment to the condition comparison register \code{x21} could change the branch outcome (W2).
For this reason, \specdoc{} discards all transient windows containing backward jumps.
As a result, existing fuzzers are limited to exploring only a restricted subset of transient windows.



\keypoint{\challenge{1-2}. Inefficient Training.}
Making the fuzzer recognize the microarchitecture changes caused by randomly generated instructions and subsequently exploit them is exceptionally challenging.
\introspec{} and \teesec{} use a manual software execution model to assist in setting up the required microarchitecture but cannot train states beyond the model.
\specdoc{} also has difficulty assembling matched training-exploitation instruction pairs because meaningless random training instructions often occupy the required addresses.
Unutilized microarchitecture training instructions not only reduce the fuzzing throughput but also diminish the training effectiveness due to potential conflicts.


The root cause of the above challenges is the address space conflict.
Since the fuzzer cannot predict training effectiveness or transient window locations, instructions are hardly placed at the desired address.
For example, training instructions may occupy addresses needed for transient windows, and different training instructions cannot be tested at the same address.
This makes it difficult for existing fuzzers to arrange instructions linearly to trigger the desired transient execution behaviors.





\textbf{Microarchitectural Observability}~\cite{zhang2023m, hofmann2023speculation, tran-black2023hide} concerns the ability of a fuzzer to monitor and measure the effects of sensitive data on the microarchitecture.
Despite having complete access to processor internal states, existing fuzzers fail to track how sensitive data propagates through the microarchitecture, leading to two challenges.

\keypoint{\challenge{2-1}. Feedback Gap.}
Prior work ignores the coverage matrix and thus fails to provide feedback for input mutation, leading to blind and random input mutation.
This problem is caused by the lack of ability to track the propagation process of sensitive data.
\introspec{} and \teesec{} cannot capture secrets after arithmetic operations due to the use of value matching.
\specdoc{} only computes the hash of the final state, and the compressed execution process prevents capturing the different propagation paths during execution.
The missing coverage matrix leaves a gap between input mutation and execution, making it difficult for the fuzzer to explore all possible transient behaviors efficiently.


\keypoint{\challenge{2-2}. Imprecise Oracle.}
Buffers are extensively used in processor microarchitecture to improve performance and typically include state registers to indicate the validity of the current data.
For example, the Line Fill Buffer (LFB) in BOOM is managed by the Miss Status Holding Register (MSHR).
Once the cache line refill is completed, MSHR switches its state register to \code{invalid} to indicate that the data in the LFB is outdated instead of clearing the LFB.
Existing work has incorrectly considered this scenario as vulnerable, as \introspec{} and \teesec{} would match the sensitive data remaining in the LFB.
It would also cause \specdoc{} to generate different hashes.
Due to the imprecise oracles, existing fuzzers pass these false positives to subsequent steps, resulting in meaningless execution.


The root cause of the above challenges is the lack of a mechanism to track state changes caused by sensitive data.
Without the ability to observe the information flow of sensitive data, existing fuzzers are unable to measure coverage based on the distribution of encoded sensitive data or query state registers to identify exploitable leakages.

\subsection{Dynamic Swappable Memory}
\label{sec:swapmem}

\begin{figure}[t]
    \centering
    \includegraphics[width=0.83\linewidth]{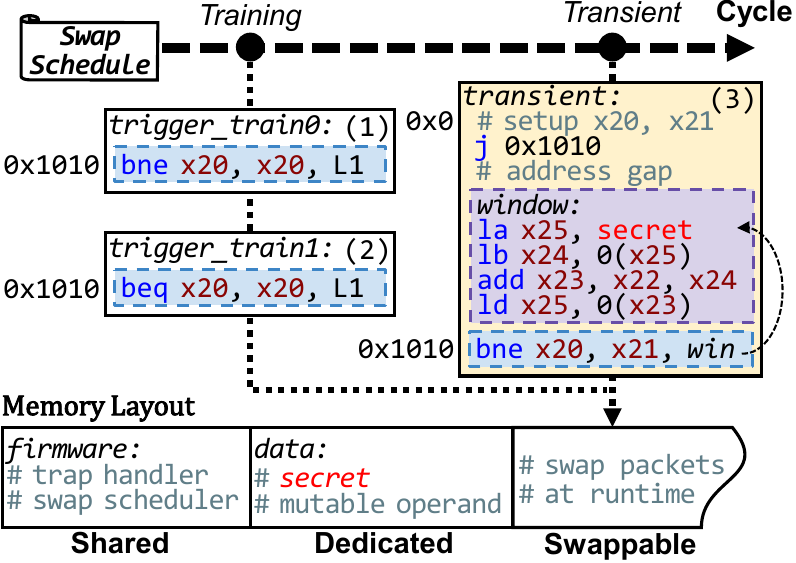}
    \caption{Using \swapmem{} to trigger transient execution.}
    \label{fig:swap-mem}
\end{figure}

Instead of using scalability-limited templates to solve the address space conflict, the core insight of \name{} is that address space can be time-shared by different semantics.
\cref{fig:swap-mem} shows how scheduling instruction sequences within the same address space enables triggering complex transient windows that could not be generated in \cref{fig:conflict}.
During simulation, we first load training instruction sequence (1) or (2) into memory to train the predictor.
After training, we flush the memory and load transient instruction sequence (3) to trigger the backward transient window at \code{0x1010}.
For the training instruction sequence, since the full address space is available, we do not need to use similar addresses like \code{0x0010} to train the predictor.
Instead, we can directly place a branch training instruction at \code{0x1010}.
Additionally, we can explore different training effects, such as using sequence (1) to train the prediction as untaken or sequence (2) to train it as taken.
For the transient instruction sequence, since training instructions are not in this sequence, W1 type conflicts are avoided, and conflicted register assignments can be moved to other available addresses (e.g., \code{0x0}) to resolve W2 type conflicts.
As shown in sequence (3), after setting up the registers, DUT can directly jump to \code{0x1010} to trigger the transient window without any conflicts.
Besides generating arbitrary transient windows, we can also identify effective training by trying different training instruction sequences.
For example, by trying combinations (1)(3) and (2)(3), we can find that only (2) contributes to triggering the transient window.
Thus, switching instruction sequences on demand at different stages effectively resolves address space conflicts, allowing the fuzzer to effectively control the microarchitecture to trigger desired transient execution behaviors.

However, implementing the above switching process with assembly instructions can pollute memory-related training states.
To address this, we propose the dynamic swappable memory (\swapmem{}), enabling transparent instruction sequence switching.
Since side channel bugs require multiple DUT instances with different secrets to detect behavioral differences, the \swapmem{} is specifically designed for this scenario.
As shown at the bottom of \cref{fig:swap-mem}, the \swapmem{} consists of three regions.
The shared region is shared across multiple DUT instances and contains the essential execution environment, including state initialization, trap handling, and runtime instruction sequence scheduling.
To facilitate modifying secrets, each DUT has a dedicated region for storing sensitive data and mutable operands.
The swappable region is used to hold instruction sequences with different semantics.
Each DUT can load the required instruction sequence into the swappable region at runtime according to the swap schedule.
Typically, \name{} first executes all training instruction sequences on the DUT, then updates sensitive data permissions, and finally executes the transient instruction sequence.
Once a sequence is completed, an exception is triggered, and then the trap handler flushes the instruction cache and loads the next sequence into the swappable region.
After swapping the new sequence, the DUT jumps to its entry and continues execution.

The \swapmem{} enhances microarchitectural controllability as the isolation primitive, resolving address space conflicts.
In \cref{sec:step1}, we will discuss how to design instruction sequence generation strategies based on \swapmem{} to trigger diverse windows and optimize training overhead.

\subsection{Differential Information Flow Tracking}
\label{sec:diffift}

\name{} intends to employ the information flow tracking technique to identify state changes caused by secrets.
However, as discussed in \cref{sec:ift}, the control flow over-tainting problem makes it impossible to identify the propagation of sensitive data.
Thus, we propose differential information flow tracking (\diffift{}) to mitigate the over-tainting problem.

When fuzzing transient execution vulnerabilities, we can assume that leakage occurs when executing a given instruction sequence using different secrets produces different behaviors.
However, \cref{rule:mux} considers arbitrary input differences rather than differences caused by secrets.
Therefore, a core insight of \name{} is that if no secret can influence the value of a control signal, then even if it is tainted, it should be ignored, as it cannot select an alternative path.
However, it is extremely expensive to precisely compute all potential values of each control signal in the out-of-order processor for all input secrets at each cycle~\cite{ifttheory}.
Inspired by the multi-variant execution~\cite{cox2006n, salamat2009orchestra, osterlund2019kmvx}, \name{} approximates the solution with concrete values from multiple variants.
To be specific, \name{} creates a differential testing testbench to determine if sensitive data can produce different values of a control signal by executing the same instructions on two identical DUTs with different secrets.
\cref{tab:diva-rule} lists the updated control taint propagation rules for all supported control flow cells.
The overall policies are similar to \cellift{}, except the control taints only propagate when cross-instance comparison signals are high.
The highlighted signals with the $\mathit{diff}$ subscript represent cross-instance comparison signals.
Take the multiplexer as an example, when \diffift{} encounters a multiplexer whose selection signal $S$ is tainted, \diffift{} checks whether the selection signals are consistent between the variants (i.e., $\diffsig{S} = S_{DUT_1} \textasciicircum S_{DUT_2}$).
If there is a difference, it indicates that sensitive data can generate different selections, and \diffift{}, therefore, performs control taint propagation.
Otherwise, \diffift{} only considers data taint propagation.
We instrument the DUT at the RTL IR level and thus support word-level cells and non-flattened memories.
Additionally, the data taint propagation policies for data flow cells in \diffift{} are consistent with \cellift{}.

\begin{table}[t]
    \centering
    \caption{The control taint propagation policies of \diffift{}.}
    \resizebox{0.95\linewidth}{!}{
    \begin{tabular}{c|c}
        \hline
        Cell Type & Propagation Policy \\ \hline
        Multiplexer       & $(\trippleop{S}{B^{t}}{A^{t}}) | (\trippleop{S^{t}\policysym{\&}\diffsig{S}}{(A \textasciicircum B)|(A^{t} | B^{t})}{0})$ \\ \hline
        Comparison Cell   & $\diffsig{O}\;\&\;|(A^t|B^t)$ \\ \hline
        Register with En & $(\trippleop{En}{D^{t}}{Q^{t}}) | (\trippleop{En^{t}\policysym{\&}\diffsig{En}}{(D \textasciicircum Q)|(D^{t} | Q^{t})}{0})$ \\ \hline
        Memory Read      & $mem^t[addr]|\{\mathit{WIDTH}\{\diffsig{addr}\}\}$ \\ \hline
        \multirow{2}{*}{Memory Write} & 
            \multirow{2}{*}{
                \begin{tabular}[c]{@{}c@{}}
                    $(\trippleop{\mathit{Wen}}{\mathit{Wdata}^{t}}{mem^{t}[addr]})|$ \qquad\qquad\qquad\qquad \\ 
                    \qquad\qquad$\{\mathit{WIDTH}\{\diffsig{\mathit{Wen}}|(\diffsig{addr}\policysym{\&}\mathit{Wen})\}\}$
                \end{tabular}} \\ 
        & \\ \hline
    \end{tabular}
    }
    \label{tab:diva-rule}
\end{table}

It is worth noting that \diffift{} is an underapproximation of information flow since it uses concrete values.
If a secret pair happens to produce the same value on a secret-dependent control signal, a false negative will occur.
When this happens, data taints still propagate accurately, but control taints are suppressed due to identical control signals.
Therefore, \name{} generates secrets for the variant DUT by flipping each bit of the original secret to avoid using identical values.
Besides, by leveraging the dedicated region in \swapmem{}, \name{} can directly load different secret pairs to mitigate false negatives without regenerating the input.

The \diffift{} serves as the tracing primitive to enhance microarchitectural observability.
With the help of taints, \name{} is able to observe sensitive data and its derived values across the microarchitecture.
In \cref{sec:step2} and \cref{sec:step3}, we will explain how to use taint to compute coverage and identify leakages.

\section{The \name{} Framework}
\label{sec:design}

\begin{figure*}[t]
    \centering
    \includegraphics[width=0.95\linewidth]{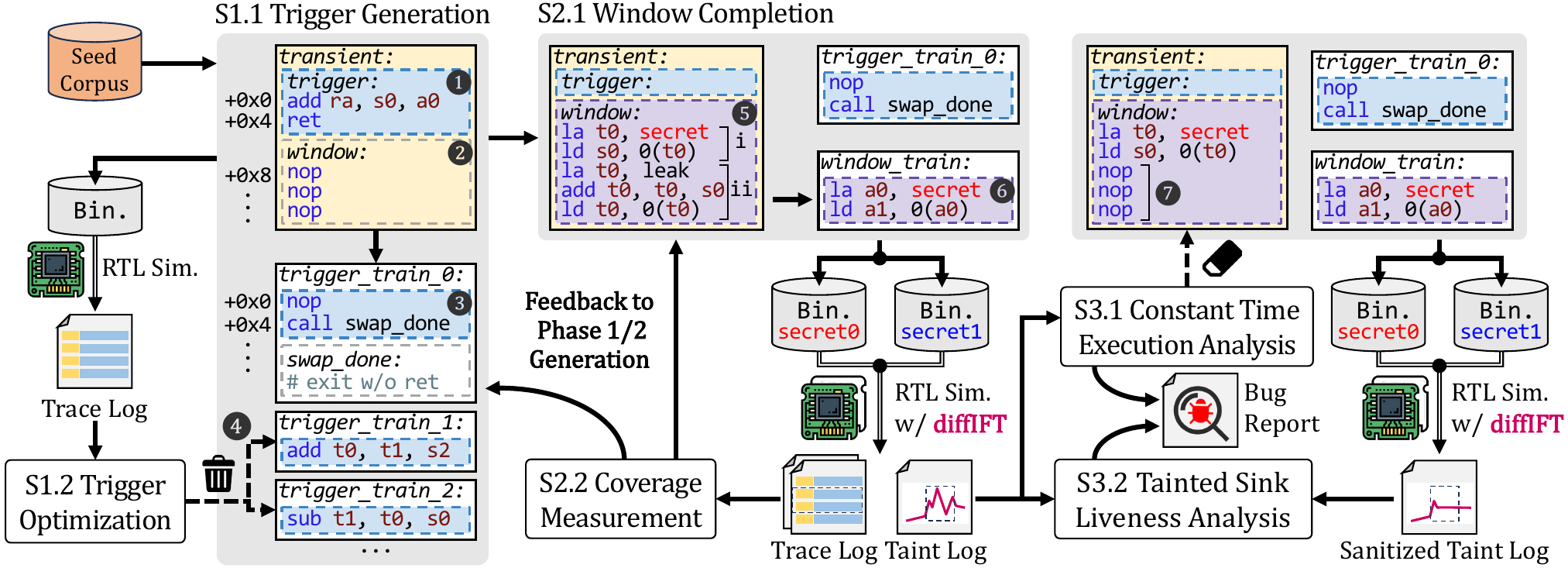}
    \caption{\name{} fuzzing workflow for finding transient execution vulnerabilities, taking Spectre-RSB for example.}
    \label{fig:fuzz-flow}
\end{figure*}


In this section, we demonstrate how \name{} builds on operating primitives to address the challenges in \cref{sec:challenge}, enabling effective and efficient transient execution bug fuzzing.


\keypoint{Overview.}
As shown in \cref{fig:fuzz-flow}, the workflow of \name{} consists of three phases.
The first two phases focus on triggering and exploring transient execution, while the final phase is responsible for detecting leakage.
\name{} leverages \swapmem{} to isolate different instruction sequences within the same address space.
In Phase 1, \name{} derives targeted training for diverse transient windows and evaluates each training to eliminate ineffective training.
In Phase 2, \name{} completes the transient window and attempts to encode sensitive data into the microarchitecture.
During simulation, \name{} uses \diffift{} to track sensitive data propagation and collects taint as coverage to guide exploration.
In Phase 3, \name{} first checks transient window constant time execution violations.
If no timing differences are detected, it further uses taint liveness annotations to check whether secrets encoded into the microarchitecture can be exploited.
Finally, \name{} reports test cases that violate transient window constant time execution or contain exploitable taints as potential bugs.


\subsection{Phase 1: Transient Window Triggering}
\label{sec:step1}

Phase 1 focuses on triggering diverse transient windows with minimal overhead.
For challenge \challenge{1-1}, \name{} uses \swapmem{} to isolate transient execution from training to generate arbitrary transient windows, and employs the training derivation strategy (\cref{sec:trig-gen}) to generate targeted training.
For challenge \challenge{1-2}, \name{} further isolates each training and applies the training reduction strategy (\cref{sec:trig-opt}) to identify and eliminate ineffective training.



\subsubsection{Step 1.1: Trigger Generation}
\label{sec:trig-gen}
While \swapmem{} resolves address space conflicts, allowing \name{} to generate arbitrary transient windows, effective training is still required to trigger them.
To train the required microarchitecture components for triggering transient windows, \name{} employs the training derivation strategy.
It first randomly generates a transient window and then derives targeted training based on the expected transient window.

\keypoint{Trigger Instruction Generation.}
In this step, \name{} only generates the \code{trigger} section of the transient packet (\bcircled{1}).
The transient packet refers to the instruction sequence that triggers a transient window and transiently accesses and encodes sensitive data (i.e., transient instruction sequence (3) in \cref{fig:swap-mem}).
\name{} first randomly generates trigger instructions based on the trigger type from the seed.
The trigger instructions supported by \name{} cover the entire basic instruction set, including sequential execution instructions (e.g., integer or floating-point arithmetic operations, valid memory accesses), control transfer instructions (e.g., branches, indirect jumps, and returns), and instructions that may trigger architectural exceptions (e.g., illegal instructions, memory access violations).
In the example shown in \cref{fig:fuzz-flow}, suppose \name{} plans to trigger a transient window caused by a return address misprediction.
Next, \name{} generates a dummy transient window filled with \code{nop} instructions (\bcircled{2}).
For sequential execution instructions and exceptions, the transient window is placed immediately after the trigger instruction by default.
For control transfer instructions, \name{} randomly selects whether to place the transient window after the trigger instruction.
Finally, \name{} uses an ISA simulator to compute the operands required to trigger the transient window and generate the related register initialization instructions.
Therefore, \name{} covers transient windows triggered by all instruction types, effectively enhancing transient window diversity.



\keypoint{Trigger Training Derivation.}
\name{} uses the transient execution information in transient packets to randomly generate multiple trigger training packets (\bcircled{3}).
The trigger training packet refers to the instruction sequence used for training microarchitecture to trigger the transient window (i.e., training instruction sequences (1) and (2) in \cref{fig:swap-mem}).
For each trigger training packet, \name{} first generates a random training instruction, and then inserts \code{nop} instructions to align it with the trigger instruction in the transient packet.
In the example, we generate three trigger training packets, with the training instructions all placed at the same address (i.e., \code{0x4}) as the trigger instruction \code{ret}.
Next, \name{} further adjusts the control flow of the training instruction if the training instruction is a control transfer instruction.
To be specific, \name{} adjusts the control flow of the training instruction to match the control flow of the generated transient window, enhancing the training effectiveness for control flow prediction.
For example, \name{} adjusts the caller address in packet \code{trigger_train_0} to ensure that the return address matches the start address of the transient window (i.e., \code{0x8}).
By deriving training from transient execution information, \name{} not only generates diverse transient windows but also produces targeted training, ensuring the fuzzer can more effectively control the microarchitecture to trigger desired transient execution behaviors.


\subsubsection{Step 1.2: Trigger Optimization}
\label{sec:trig-opt}
After generating the trigger training packets, \name{} evaluates which packets are helpful in triggering transient windows.
Leveraging \swapmem{}, \name{} employs the training reduction strategy that identifies and discards ineffective trigger training packets without affecting transient window triggering, thereby reducing training overhead.

\keypoint{Transient Execution Evaluation.}
\name{} packages these packets together with a swap schedule, which schedules them in the order of trigger training packets first and then the transient packet.
After the RTL simulation, \name{} analyzes the RoB IO events from the trace log.
If the number of enqueued instructions within the transient window exceeds the number of its committed instructions, it indicates that the transient window has been successfully triggered.

\keypoint{Training Reduction.}
Although trigger training packets are derived from the transient packet for targeted training, not all training contributes to triggering the transient window.
Fortunately, since each training instruction is isolated in its packet, \name{} can identify ineffective packets by removing one at a time and re-simulating the remaining packets to see if the transient window still triggers (\bcircled{4}).
If removing a trigger training packet does not affect transient window triggering, it will be permanently discarded from the swap schedule.
Otherwise, the packet is necessary, and \name{} will keep it in the swap schedule.
\name{} evaluates each trigger training packet in the order of the swap schedule.
This process repeats until only necessary trigger training packets remain or none are available.
It is obvious that integer arithmetic operations do not contribute to return address prediction.
Therefore, in the example, \name{} finds that discarding \code{trigger_train_1} and \code{trigger_train_2} does not affect the triggering of the transient window, and finally removes them.
By discarding ineffective trigger training packets, \name{} is able to trigger transient windows with minimal training overhead.




\subsection{Phase 2: Transient Execution Exploration}
\label{sec:step2}

\name{} explores which microarchitectures can be used to encode secrets during this phase.
\name{} uses taints as the coverage to guide the exploration (\cref{sec:coverage}), effectively addressing challenge \challenge{2-1}.


\subsubsection{Step 2.1: Window Completion}

\name{} replaces the dummy transient window with real payloads and generates a complete test case.

\keypoint{Transient Window Completion.}
\name{} generates two blocks in the \code{window} section (\bcircled{5}): (i) the secret access block and (ii) the secret encoding block.
In the secret access block, besides fixed instructions to access sensitive data, it also randomly masks the high-order bits of the address to attempt to cover MDS-type bugs.
In the secret encoding block, \name{} randomly generates instructions that depend on secrets in order to propagate secrets across the microarchitecture.

\keypoint{Window Training Derivation.}
Similar to trigger training packets, \name{} also derives window training packets for the secret access block (\bcircled{6}).
The window training packet is the training instruction sequence used to train memory-related states used by the transient window.
In the example, \name{} attempts to warm up sensitive data into the processor's internal buffers in advance, such as data cache and load buffer.
The generated window training packets are scheduled before the trigger training packets in the swap schedule to avoid invalidating the transient window.

\subsubsection{Step 2.2: Coverage Measurement}
\label{sec:coverage}

\name{} performs RTL simulation using the \diffift{} instrumented DUTs and measures coverage from the taint log to guide subsequent stimulus generation.

\keypoint{Taint Coverage.}
\name{} introduces the first secret sensitive coverage matrix designed for transient execution vulnerability fuzzing.
The taint coverage treats the total number of taints within a local range as an independent coverage point.
To be specific, \name{} inserts a new register array \code{bitmap} into each RTL module.
During each clock cycle, \name{} uses the number of tainted registers within the module as the index and writes 1 to the corresponding slot in the \code{bitmap}.
After the transient execution, \name{} checks the value of each slot in the \code{bitmap}.
If a slot's value is 1, it indicates that the corresponding number of taints has been explored within the module, and \name{} records the index of such a slot and its module name as a tuple.
Finally, \name{} evaluates input exploration based on the total number of collected \code{(module, index)} tuples.

The taint coverage has two key properties.
The first is locality, as coverage is measured at the module level, reflecting the propagation of sensitive data across different hierarchies.
The second is position-insensitive, which helps filter out redundant encoding.
For example, when sensitive data is encoded in different slots of the cache data array, the coverage points generated by the cache module are the same.

\keypoint{Coverage Feedback.}
Once all packets are ready, \name{} packages them into two swappable stimuli with different secrets for \diffift{}.
After simulation, \name{} first determines the cycle range of the transient window by analyzing RoB IO events from the trace log, then checks taint changes in the transient window from the taint log.
If taints increase, it indicates that sensitive data has been successfully propagated, and \name{} further measures the taint coverage from the taint log.
If the coverage increase is less than the average increase or sensitive data is not propagated, \name{} mutates the seed to regenerate the \code{window} section.
If the results after multiple attempts still show low coverage growth, \name{} will discard the seed and return to Phase 1.

\subsection{Phase 3: Transient Leakage Analysis}
\label{sec:step3}


In this phase, \name{} analyzes whether the final state can leak sensitive data.
For challenge \challenge{2-2}, \name{} uses taint liveness annotations to filter out unexploitable taints in the final analysis phase (\cref{sec:liveness}).

\subsubsection{Step 3.1: Constant Time Execution Analysis}
For test cases that successfully access and propagate sensitive data, \name{} further analyzes whether leakage occurred.
It first compares the execution time of the transient window between DUTs.
If inconsistent, it indicates that sensitive data may have caused timing side channels, such as port contention, during the transient window.
\name{} directly reports these test cases as potential vulnerabilities.

\keypoint{Encode Sanitization.}
Although test cases with transient window constant time execution cannot directly leak secrets through the timing side channel, the encoded sensitive data may still be leaked via other side channels.
Since accessing sensitive data during training also generates taints, we need to distinguish the taints caused by the secret encoding block before further analyzing whether the encoded sensitive data can be exploited.
Therefore, \name{} replaces the secret encoding block in the transient packet with \code{nop} instructions (\bcircled{7}) and re-runs the simulation.
By comparing the sanitized taint log with the original taint log, \name{} can identify the taints generated by the secret encoding block.


\subsubsection{Step 3.2: Tainted Sink Liveness Analysis}
\label{sec:liveness}

The taints produced by \diffift{} only indicate reachability.
As the LFB example in \cref{sec:challenge}, not all encoded secrets are exploitable.
Therefore, \name{} further analyzes taint liveness to determine whether the tainted sinks can be exploited.

\keypoint{Taint Liveness Annotation.}
Inspired by selective data protection~\cite{regvault, dynpta, ahmed2023not}, \name{} uses annotations to bind taint registers to their corresponding state registers.
Developers can annotate the registers with the \code{liveness_mask} custom attribute~\cite{thomas2008verilog, design2009ieee} to declare their state registers.
Taking LFB as an example, the \code{mshr_valid_vec} signal comes from the state register in MSHR, and the \code{lb} register is the data buffer in LFB.
Line 4 shows the annotation.
During \diffift{} instrumentation, \name{} automatically connects the liveness signal \code{mshr_valid_vec} to the taint register of \code{lb}.

\begin{lstlisting}[style={verilog-style}]
wire mshrs_0_valid, mshrs_1_valid;
wire [15:0] mshr_valid_vec = 
                {8{mshrs_1_valid}, 8{mshrs_0_valid}};
(* liveness_mask = "mshr_valid_vec" *)
reg [63:0] lb [15:0];

BoomMSHR mshrs_0 (.io_mshr_valid(mshrs_0_valid));
BoomMSHR mshrs_1 (.io_mshr_valid(mshrs_1_valid));
\end{lstlisting}

However, since the implementation of the state registers is coupled with the microarchitecture, developers may be unable to reference them directly.
To accommodate various implementation, we design the liveness signal interface as a generic vector, with each bit representing whether the corresponding slot in the taint register array is valid.
For example, the lower 8 entries of \code{lb} are managed by \code{mshrs_0}, while the upper 8 entries are managed by \code{mshrs_1}.
We can construct the liveness signal as shown in lines 2-3.
\name{} currently requires developers to manually convert state registers into liveness signal vectors.
\cref{tab:dut} shows the manual effort required for annotation and patching.
By default, \name{} treats all register arrays (including those registers generated by \code{Vec} in Chisel) as potential sinks, and developers can customize sinks as needed.
Finally, \name{} identifies the target sinks from the encoded secrets obtained in the previous step and reports tainted sinks with valid liveness signals as potential vulnerabilities.
\section{Implementation}

The implementation consists of 1) a testharness generator responsible for instrumenting RTL source code and integrating two DUTs into a testbench containing \swapmem{}, and 2) the fuzzing pipeline illustrated in \cref{fig:fuzz-flow}.

\keypoint{Testharness Generator.}
We implement the \swapmem{} atop the Starship SoC generator~\cite{starship}, with $\sim$300 LoC Python for \swapmem{} RTL model generation and $\sim$500 LoC DPI-C for \swapmem{} runtime.
The \diffift{} instrumentation adds new passes in the Yosys synthesizer to insert taint cells for taint propagation, involving $\sim$1KLoC C++.
The taint cell library of \diffift{} is implemented in Verilog, which also uses $\sim$1KLoC.


\keypoint{Fuzzing Pipeline.}
The fuzzing pipeline consists of $\sim$6500 LoC Python and $\sim$180 LoC RISC-V assemble, which includes stimulus generation and fuzzing management.
\name{} uses seeds to generate stimuli, which contain configurations for trigger instructions and transient windows, as well as entropy for the random instruction generator.
The generator supports the RV64GC instruction set and covers common transient window types. 
The fuzzing manager employs a multi-threaded design, allowing multiple RTL simulation instances to run in parallel.

\begin{table}[t!]
    \centering
    \caption{Summary of the cores used for evaluation.}
    \label{tab:dut}
    \resizebox{0.8\linewidth}{!}{
    \begin{tabular}{lcc}
        \hline
        Feature & BOOM & XiangShan \\
        \hline
        Configuration & SmallBOOM & MinimalConfig \\
        ISA & RV64GC & RV64GC \\
        Verilog LoC & 171K & 893K \\
        Annotation LoC & 212 & 592 \\
        \hline
        \end{tabular}
    }
\end{table}

\section{Evaluation}

\begin{table*}[ht!]
    \centering
    \caption{Training overhead for different types of transient windows.}
    \label{tab:trigger-eval}
    \resizebox{0.95\textwidth}{!}{
        \begin{threeparttable}
        \begin{tabular}{ccc|c|c|c|c|c|c|c}
        \hline
            \multirow{2}{*}{Processor} &
            \multirow{2}{*}{Fuzzer} &
            \multicolumn{1}{c|}{\begin{tabular}[c]{@{}c@{}}Load/Store\\ Access Fault\end{tabular}} &
            \multicolumn{1}{c|}{\begin{tabular}[c]{@{}c@{}}Load/Store\\ Page Fault\end{tabular}} &
            \multicolumn{1}{c|}{\begin{tabular}[c]{@{}c@{}}Load/Store\\ Misalign\end{tabular}} &
            \multicolumn{1}{c|}{\begin{tabular}[c]{@{}c@{}}Illegal\\ Instruction\end{tabular}} &
            \multicolumn{1}{c|}{\begin{tabular}[c]{@{}c@{}}Memory\\ Disambiguation\end{tabular}} &
            \multicolumn{1}{c|}{\begin{tabular}[c]{@{}c@{}}Branch\\ Misprediction\end{tabular}} &
            \multicolumn{1}{c|}{\begin{tabular}[c]{@{}c@{}}Indirect Jump \\ Misprediction\end{tabular}} &
            \multicolumn{1}{c}{\begin{tabular}[c]{@{}c@{}}Return Address\\ Misprediction\end{tabular}} \\ \cline{3-10}
            & &  TO (ETO) &  TO (ETO) &  TO (ETO) &  TO (ETO) &  TO (ETO) & 
                 TO (ETO) &  TO (ETO) &  TO (ETO) \\ \hline
            \multirow{3}{*}{BOOM} & 
              \name{}     &  0.0 (0.0)  &  0.0 (0.0) &  0.0 (0.0)  &  \faTimes{} &  0.0 (0.0) & 86.4 (3.8) &  85.7 (2.8) &  85.6 (2.7) \\
            & \name{}$^{*}$ &  1.3 &  0.1 &  1.6 &  \faTimes{} &  0.2 & 102.2 & 169.5 &  89.5 \\
            & \specdoc{} &  \faTimes{} &  126.6 &  \faTimes{} &  \faTimes{} &  113.5 &  125.5 &  122.5 &  \faTimes{} \\ \hline
            \multirow{2}{*}{XiangShan} &
              \name{}     &  0.1 (0.0) &  0.0 (0.0) &  0.0 (0.0) &  0.0 (0.0) &  0.0 (0.0) & 83.9 (2.8) &  90.1 (2.9) &  88.7 (2.9) \\
            & \name{}$^{*}$ &  0.0 &  0.0 &  0.0 &  0.0 &  0.4 & 101.0 &  \faTimes{}  &  97.0 \\ \hline
        \end{tabular}
            \begin{tablenotes}
                \item \faTimes{} indicates that the corresponding type of transient window failed to trigger.
            \end{tablenotes}
        \end{threeparttable}
        }
\end{table*}

We evaluate \name{} by answering the following questions:

\begin{itemize}[noitemsep, leftmargin=*]
\item \keypoint{RQ 1.} {How effective and efficient is \name{} in triggering diverse transient windows?} (\cref{eval:control})

\item \keypoint{RQ 2.} {How well does \name{} trace sensitive data, improve coverage, and identify leakages?} (\cref{eval:observe}) 

\item \keypoint{RQ 3.} {Can \name{} uncover previously unknown transient execution bugs in real-world processors?} (\cref{eval:bug})
\end{itemize}

\subsection{Experimental Setup}

All experiments are conducted on a machine with dual AMD EPYC 9334 processors featuring 64 cores and 512GB of RAM.
We use the industry-standard RTL simulator Synopsys VCS for RTL simulation.
Limited by the number of licenses, we only used a maximum of 16 threads in the experiments.

We evaluate \name{} on BOOM~\cite{boom} and XiangShan~\cite{xiangshan}, two well-known out-of-order processors that are actively maintained in the RISC-V community.
BOOM is the third generation of the Berkeley out-of-order machine and is widely evaluated in related academic work~\cite{introspectre, teesec, specdoctor, fadiheh2022exhaustive, tan2024rtl, cellift}.
XiangShan is currently the most high-performance open-source RISC-V core and thus has a more complex architecture.
Their configurations are summarized in \cref{tab:dut}.

Since \introspec{} and \teesec{} only focus on Meltdown-type vulnerabilities and their released artifacts do not include a complete fuzzing framework, we only compare \name{} with \specdoc{}.
Due to the complex manual patching of the DUT required by \specdoc{}, we only compare the BOOM supported by both.

\subsection{Microarchitectural Controllability Evaluation}
\label{eval:control}

We collect 2,500 transient windows separately and summarize their types and training overhead in \cref{tab:trigger-eval}.
The Training Overhead (TO) refers to the number of training instructions generated to trigger transient windows.
Since \name{} uses \code{nop} instructions to align training instructions with trigger instructions, we also compute the Effective Training Overhead (ETO) by excluding the padding \code{nop} instructions.
For misprediction-type transient windows, since predictors have default prediction states, we exclude transient windows that require no training to trigger.

The results show that \specdoc{} can only cover 4 types of transient windows on BOOM and requires about 125 instructions for training.
Instead, \name{} can trigger all types of transient windows with minimal overhead.
Notably, the training reduction strategy successfully identifies the necessary training packets for triggering the transient window.
Therefore, \name{} can trigger exception-type transient windows with zero overhead and use a few training instructions (excluding \code{nop} instructions) to trigger misprediction-type windows.
To show the effectiveness of the training derivation strategy, we introduce the \name{}$^*$ variant.
\name{}$^*$ still uses \swapmem{}, but its training packets consist of random instructions instead of deriving from transient execution information.
Due to the training reduction strategy, both \name{}$^*$ and \name{} have zero training overhead for exception-type transient windows.
However, since random training fails to align trigger instructions and match transient execution flows, \name{}$^*$ cannot trigger indirect jump misprediction on XiangShan.
For the other misprediction-type transient window, \name{}$^*$ incurs higher training overhead due to the lack of targeted training.
These results demonstrate that \name{} can effectively and efficiently trigger more diverse transient windows.

\begin{table}[t]
    \centering
    \caption{Overhead of differential information flow tracking.}
    \label{tab:ift}
    \resizebox{\linewidth}{!}{
        \begin{tabular}{cc|ccc|ccc}
            \hline
            \multicolumn{2}{c|}{\multirow{2}{*}{Time (s)}} & \multicolumn{3}{c|}{BOOM}    & \multicolumn{3}{c}{XiangShan} \\ \cline{3-8} 
            \multicolumn{2}{c|}{}                                         & Base & \cellift{} & diffIFT & Base & \cellift{} & diffIFT \\ \hline
            \multicolumn{2}{c|}{Compile}                                     & 122   & 2856   & 268   & 638 & \multirow{6}{*}{\begin{tabular}[c]{@{}c@{}}Timeout\\ after 8h\end{tabular}} & 1781 \\ \cline{1-6} \cline{8-8}
            \multirow{5}{*}{\rotatebox{90}{Simulation}}    & Spectre-V1      & 2.0   & 152.2  & 4.8   & 4.0 & & 17.5 \\
                                                           & Spectre-V2      & 2.1   & 152.4  & 5.7   & 4.5 & & 19.8 \\
                                                           & Meltdown        & 2.1   & 152.6  & 5.6   & 4.7 & & 19.9 \\
                                                           & Spectre-V4      & 2.0   & 152.2  & 4.9   & 4.3 & & 17.9 \\
                                                           & Spectre-RSB     & 2.0   & 152.0  & 4.8   & 4.3 & & 17.9 \\ \hline
        \end{tabular}
    }
\end{table}

\subsection{Microarchitectural Observability Evaluation}
\label{eval:observe}

\begin{figure}[t]
    \centering
    \includegraphics[width=0.95\linewidth]{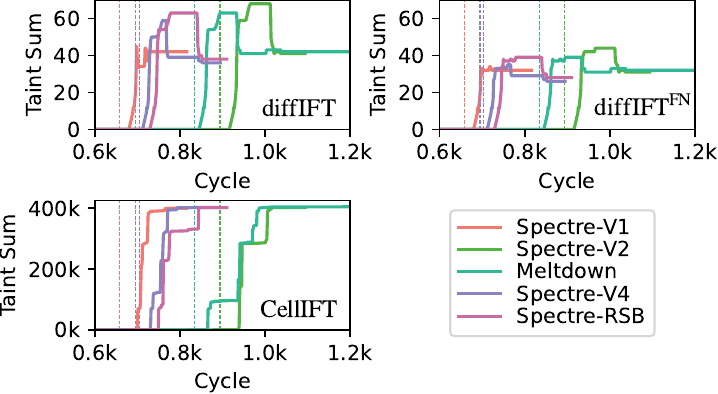}
    \caption{Taints during executing each test case. The dotted vertical line represents the start of the transient execution.}
    \label{fig:taint}
\end{figure}

\keypoint{Micro-benchmark.}
We first evaluate the overhead of \diffift{} instrumentation at compile and runtime, using the state-of-the-art information flow tracking technique \cellift{} as a reference. 
The compilation duration includes Chisel elaboration, Yosys instrumentation, and VCS synthesis.
For runtime overhead, we manually implement a benchmark covering common transient execution vulnerability test cases and record simulation times.
\cref{tab:ift} shows the results, indicating that the overhead of \diffift{} is acceptable compared to \cellift{}.
Since \cellift{} instruments at the cell level, it requires flattening all memory, resulting in a significantly increased compilation time.
In contrast, \diffift{} instruments at the RTL IR level, achieving faster instrumentation.
\cref{fig:taint} further shows the changes in the taint sum over cycles when executing the benchmark on BOOM.
The result proves that \cellift{} does suffer from taint explosion.
Once all registers are tainted, \cellift{} loses the ability to track secrets, and the simulation speed is severely degraded. 
By eliminating control taints caused by identical control signals, \diffift{} effectively mitigates control flow over-tainting.
Even with two DUTs instantiated in the testbench, the runtime overhead of \name{} is still acceptable.

And to understand the impact of false negatives, we also introduce the \diffift{}$^\mathrm{FN}$ variant in \cref{fig:taint}.
In the \diffift{}$^\mathrm{FN}$ variant, the two DUT instances in the testbench use the same secret to ensure all control signals are identical, representing the worst-case scenario of false negatives.
After the transient window is triggered, the taint gradually increases as the secret is loaded into registers.
However, since all control signals are the same, \diffift{}$^\mathrm{FN}$ fails to propagate control taints during the process of encoding sensitive data, causing the taints to stop increasing.
Finally, the remaining taints are data taints carried by residual secrets in multiple caches and buffers.
Therefore, when false negatives occur, data taints still propagate accurately, but control taints are suppressed due to identical control signals.

\begin{figure}[t]
    \centering
    \includegraphics[width=0.85\linewidth]{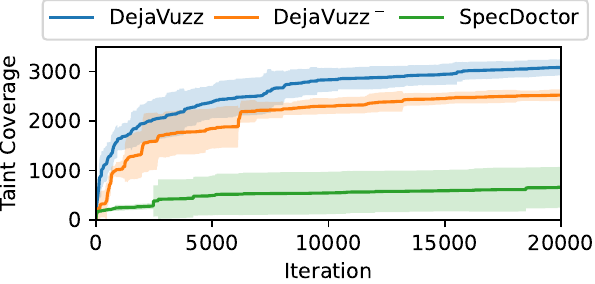}
    \caption{Taint coverage for 5 trials over 20,000 iterations.}
    \label{fig:cov}
\end{figure}

\keypoint{Coverage Evaluation.}
Next, we evaluate the efficiency of microarchitecture exploration.
\cref{fig:cov} illustrates the growth trend of taint coverage on BOOM.
Each experiment is repeated 5 times, and the shaded area represents the 95\% confidence interval.
To avoid the impact of simulation performance differences between different RTL simulators, we replay the phase 3 test cases generated by \specdoc{} in our environment to obtain comparable results and use the number of iterations as the x-axis.
The y-axis represents the number of taint coverage points defined in \cref{sec:coverage}.
Due to the lack of feedback on the sensitive data propagation process, \specdoc{} only performs random mutations on test cases that can produce different state hashes, limiting its ability to effectively guide fuzzing.
With the help of taints, \name{} can guide mutation more effectively, ultimately exploring 4.7$\times$ more coverage than \specdoc{}.
Moreover, \name{} achieves the same saturation coverage as \specdoc{} in just 118 iterations.
\name{}$^-$ is used to demonstrate the effectiveness of using \diffift{} as coverage.
Instead of using taint coverage, it randomly updates the secret encoding block or regenerates a new transient window for each round.
The result shows that \name{} achieves a 22\% coverage improvement over \name{}$^-$ and achieves the same coverage in 7,200 iterations that \name{}$^-$ requires 20,000 iterations to reach.
The coverage difference between them demonstrates that using taints as coverage enables more efficient microarchitecture exploration.

\keypoint{Liveness Evaluation.}
We also found an interesting phenomenon that \specdoc{} did not report any vulnerabilities during the coverage evaluation.
According to \specdoc{}'s design, its phase 3 identified a total of 75 test cases that could encode sensitive data into the timing components and generate different state hashes.
And in its phase 4, \specdoc{} attempts to generate random instructions to decode secrets from those timing components.
Unfortunately, \specdoc{} spent nearly a week executing 100,000 iterations without finding any vulnerabilities.
We use taint liveness annotations to analyze all 75 test cases, and find that only 17 of them are real leakages, while the rest are false positives.
Most false positives are caused by secrets that fail to be encoded into the microarchitecture but still remain in the data cache.
An exception is an invalid test case that executes the transient window during the training.
Limited by poor microarchitectural observability, \specdoc{} spends a significant amount of time futilely generating random instructions to decode unexploitable false positives.
To further validate the effectiveness of taint liveness annotations, we re-execute the test cases using a \name{} variant without taint liveness annotations.
Only 21 test cases are correctly identified, while the remaining 54 cases are misclassified due to residual invalid taints in physical registers or RoB.
This highlights the effectiveness of taint liveness annotations. 
With the help of the liveness signals, \name{} can identify exploitable leakages without resorting to inefficient and nondeterministic random decode instruction generation.

\subsection{Bugs Found in Real-World Processors}
\label{eval:bug}

Note that the coverage is only used to evaluate exploration, higher coverage does not guarantee more bugs.
Therefore, we also compared the bugs found during the liveness evaluation.
\cref{tab:bug} categorizes all transient execution vulnerabilities discovered by \name{} based on the attack type, transient window type, and exploited timing component.
In comparison, \specdoc{} can only encode sensitive data into the \code{dcache} or trigger \code{lsu} port contention.
Regarding first bug detection time, \specdoc{} takes several days, whereas \name{} detects the first bug in an average of about 10 minutes with 16 threads.
Similar to existing work~\cite{introspectre, fadiheh2022exhaustive, tan2024rtl}, \name{} can cover all trigger variations of known transient execution vulnerabilities, such as replacing the transient window triggered by a page fault in the Meltdown vulnerability with one triggered by unaligned memory access.
Additionally, \name{} discovers 5 previously undiscovered transient execution vulnerabilities.

\begin{table}[t]
    \centering
    \caption{Summary of discovered transient execution bugs.}
    \label{tab:bug}
    \resizebox{\linewidth}{!}{
\begin{threeparttable}
\begin{tabular}{cccc}
\hline
Processor                  & \begin{tabular}[c]{@{}c@{}}Attack \\ Type\end{tabular} & Transient Window\tnote{1} & \begin{tabular}[c]{@{}c@{}}Encoded Timing\tnote{2} \\ Component\end{tabular} \\ \hline
\multirow{7}{*}{BOOM}      & \multirow{3}{*}{Meltdown} & mem-excp & i/dcache, (l2)tlb, lsu \\ \cline{3-4} 
                           & & \multirow{2}{*}{\begin{tabular}[c]{@{}c@{}}mispred, \\ mem-disamb\end{tabular}} & \multirow{2}{*}{i/dcache, (l2)tlb} \\
                           & & & \\ \cline{2-4} 
                           & \multirow{3}{*}{Spectre} & mem-excp & \begin{tabular}[c]{@{}c@{}}i/dcache, (fau)btb, \\ ras, loop, lsu, fpu\end{tabular} \\ 
                           \cline{3-4} 
                           & & \begin{tabular}[c]{@{}c@{}}mispred,\\ mem-disamb\end{tabular} & \begin{tabular}[c]{@{}c@{}}i/dcache, ras, \\ loop, lsu, fpu\end{tabular} \\ 
                           \hline
\multirow{4}{*}{XiangShan} 
            & Meltdown & 
                \begin{tabular}[c]{@{}c@{}}mem-exp, \\ mispred, illegal, \\ mem-disamb\end{tabular} & i/dcache \\ 
                           \cline{2-4} 
            & Spectre & 
                \begin{tabular}[c]{@{}c@{}}mem-excp, \\ mispred, illegal, \\ mem-disamb\end{tabular} & \begin{tabular}[c]{@{}c@{}}i/dcache, \\ 
                           lsu, fpu\end{tabular} \\ \hline
\end{tabular}
    \begin{tablenotes}
        \item[1] \textbf{mem-excp}: load/store misalign, load/store access/page fault exceptions; \textbf{mispredict}: control-flow misprediction;
        \textbf{illegal}: illegal instruction exception;
        \textbf{mem-disamb}: memory disambiguate.
        \item[2] \textbf{lsu}: load unit contention; \textbf{fpu}: floating-point unit contention; \textbf{faubtb}: first level branch target buffer; \textbf{ras}: return address stack; \textbf{loop}: loop branch predictor. 
    \end{tablenotes}
\end{threeparttable} 
    }
\end{table}

\keypoint{B1. MeltDown-Sampling (CVE-2024-44594)} is a hybrid vulnerability of Meltdown and MDS on XiangShan, allowing attackers to sample controllable targets using illegal addresses within a transient window.
\name{} generates illegal addresses (e.g., \code{0x8000...80004000}) through the secret access blocks with masks.
Due to inconsistent wire widths, when the illegal address is sent to the load unit from the pipeline, the high-bit mask is implicitly truncated. Thus, attackers can sample the secret located at \code{0x80004000}.

\keypoint{B2. Phantom-RSB (CVE-2024-44591)} is a vulnerability on BOOM that allows transiently executed instructions to update RSB.
As shown in the code below, an attacker can corrupt the RSB based on sensitive data.
Although BOOM implements a mitigation that restores the Top-Of-Stack (TOS) pointer and the return address in the top entry after mispredictions (line 11), \name{} discovers that BOOM does not restore entries below the TOS pointer (line 10).
After the RSB is corrupted, the attacker can leak the secret by measuring the execution time of the \code{ret} instruction.


\begin{lstlisting}
beq a0, a0, foo  # Predicting the branch untaken, now TOS[*\rvcomment{\rightarrow}*]X
la t0, secret    # Loading secret
ld s0, 0(t0)
andi s0, s0, 0x1 # If secret=1, ra=addr of line6, a valid[*$\textcolor{green!50!black}{_{\rotatebox[origin=c]{180}{$\Rsh$}}}$*]
sub s0, x0, s0   # addr; else ra=0, an illegal addr
auipc ra, 0      # Following code requires ra has a valid[*$\textcolor{green!50!black}{_{\rotatebox[origin=c]{180}{$\Rsh$}}}$*]
and ra, ra, s0   # addr, illegal addr will be blocked
jalr x0, 12(ra)  # Return to next, TOS[*\rvcomment{\rightarrow}*]X-1
jalr x0, 16(ra)  # Return to next, TOS[*\rvcomment{\rightarrow}*]X-2
jalr ra, 20(ra)  # Call to next, overwrite X-1
jalr ra, 24(ra)  # Call to next, overwrite X
\end{lstlisting}

\keypoint{B3. Phantom-BTB (CVE-2024-44590)} is a vulnerability similar to \textit{Boombard}~\cite{specdoctor}, where BOOM updates the BTB for exceptions under certain conditions.
The following code illustrates the details. Due to a race condition bug in BOOM, when an indirect jump misprediction coincides with an exception commit, BOOM misinterprets the exception as an indirect jump and uses the prediction correction for the mispredicted indirect jump (line 12) to update the BTB entry (line 1) of the instruction that triggered the exception.

\begin{lstlisting}
lw t0, 1(x0)     # Triggering a misalign exception
la t0, secret    # Loading secret
ld s0, 0(t0)
andi s0, s0, 0x1 # If secret=1, ra=addr of line6, a valid[*$\textcolor{green!50!black}{_{\rotatebox[origin=c]{180}{$\Rsh$}}}$*]
sub s0, x0, s0   # addr; else ra=0, an illegal addr
auipc ra, 0      # Following code requires ra has a valid[*$\textcolor{green!50!black}{_{\rotatebox[origin=c]{180}{$\Rsh$}}}$*]
and ra, ra, s0   # addr, illegal addr will be blocked
jalr x0, 12(ra) 
nop              # Padding nop to make the final[*$\textcolor{green!50!black}{_{\rotatebox[origin=c]{180}{$\Rsh$}}}$*]
# ...            # misprediction commit with the[*$\textcolor{green!50!black}{_{\rotatebox[origin=c]{180}{$\Rsh$}}}$*]
nop              # exception in the same cycle
jalr x0, 12(ra)  # Misprediction
\end{lstlisting}

\keypoint{B4. Spectre-Refetch (CVE-2024-44592, CVE-2024-44593)} is a variant of Spectre-Rewind~\cite{fustos2020spectrerewind} discovered on both BOOM and XiangShan.
\name{} found that the instruction address can also be a resource to cause port contention.
Specifically, placing the secret dependent branch at an address that triggers instruction cache miss causes the processor to preempt the fetch component during transient execution.
This allows attackers to infer the secret by measuring the execution time of the first instruction after the transient window.

\keypoint{B5. Spectre-Reload (CVE-2024-44595)} is another variant of Spectre-Rewind on XiangShan.
\name{} found that the load pipeline and load queue contend on the load write-back port of the memory access component.
By replacing the floating-point division instructions in the secret-dependent branch of Spectre-Rewind with cache-hitting load instructions, attackers can detect increased latency in cache-missing loads before the transient window.

All of the above vulnerabilities can be exploited to leak sensitive data.
B1 can directly leak secrets across privilege boundaries, while B2–B5 require access permission for sensitive data to trigger.
We disclosed identified bugs by sending bug reports to respective communities in accordance with the security policies listed for the associated project. 
According to the maintainers, all vulnerabilities in XiangShan have been fixed, while bugs in BOOM will be retained for future research.
Therefore, we recommend against using the BOOM processor in security-critical environments.


\section{Discussion and Limitation}

\keypoint{Precision Trade-off.}
Implementing precise IFT is inevitably expensive since it is an NP-complete problem~\cite{iftnp}.
Although \diffift{} can mitigate false positives caused by control flow over-tainting, it also introduces false negatives due to the inability to exhaustively compare all secrets.
In practice, \name{}, as a dynamic verification solution, can mitigate false negatives by repeatedly attempting different secret pairs.


\keypoint{Training Preference.}
Some predictors may require longer training patterns.
For instance, in the case of branch mispredictions triggered by branch instructions, training a loop predictor to trigger requires a much longer training instruction sequence compared to training a local branch history table to trigger.
Therefore, due to the training reduction strategy, \name{} prefers to choose the least costly training instruction sequence.

\keypoint{Stimulus Migration.}
The stimuli generated by \name{} only work on \swapmem{}.
Fortunately, developers usually only need simulation waveform files to pinpoint bugs.
If the stimuli must be migrated to a standard memory model (e.g., for writing general-purpose exploitations), careful manual stitching of the packets is required. 

\keypoint{Manual Annotation.}
Since the state registers are coupled to the implementation, they and their bound taints may reside in different pipeline stages or even across modules.
Limited by the loss of semantic information during the design synthesis to RTL, \name{} currently relies on manual taint liveness annotations.
We leave the automatic taint liveness annotation (such as using type-safe hardware description languages or large language models) for future work.

\section{Related Work}

\keypoint{Processor Fuzzing.}
Encouraged by the promising results of processor fuzzing on functional bugs~\cite{difuzzrtl, thehuzz, morfuzz, cascade}, several approaches have applied processor fuzzing to transient execution vulnerabilities.
\introspec{}~\cite{introspectre} and \teesec{}~\cite{teesec} use manually crafted gadgets to generate Meltdown-type vulnerabilities and detect leakages by analyzing processor runtime logs.
\specdoc{}~\cite{specdoctor} generates stimuli for transient execution attacks in multiple phases and determines bugs by observing the final execution time.
However, these approaches have the following main limitations.
First, they linearly generate transient windows or randomly combine instructions for training, resulting in limited diversity and efficiency in triggering transient windows.
Second, they can only analyze shallow information from the microarchitecture, making it impossible to provide feedback on the propagation of sensitive data or identify exploitable leakages.
To solve these limitations, \name{} uses \swapmem{} to generate and optimize training instructions to trigger diverse transient windows efficiently, and it employs differential information flow tracking to trace sensitive data to provide coverage feedback and detect exploitable leakages.

\keypoint{Black-box Microarchitecture Fuzzing.}
Commercial processors lack interfaces for obtaining fine-grained internal state information, leading to limited fuzzing exploration space.
Most of the existing black-box fuzzers, such as SpeechMiner~\cite{yuan2020speechminer} and Transynther~\cite{moghimi2020medusa}, rely on domain knowledge and can only detect vulnerability variants within a limited template scope.
Revizor~\cite{oleksenko2022revizor, tran-black2023hide} introduces the model-based relational testing approach that generates random instructions to trigger contract violations.
However, due to the limited microarchitectural controllability, they cannot even cover some known vulnerabilities that require simple training.
Integrating \swapmem{} (e.g., through DMA) can provide better control over the microarchitecture, facilitating deeper testing of black-box processors.

\keypoint{Formal Verification.}
By rigorously defining speculative contracts~\cite{guarnieri2021hardware}, ideally, formal verification can catch all transient execution bugs or prove security.
However, in practice, today's formal verification tools usually suffer from limited scalability and cannot be directly applied to complex out-of-order processors.
To bypass this limitation, optimized verification schemes~\cite{trippel2018checkmate, wang2023specification, tan2024rtl, fadiheh2022exhaustive} verify abstract models of out-of-order processors.
However, the efficacy of such formal checks depends on the precision of the models (e.g., both B2-B4 escape previous formal analyses on BOOM).
\name{} can be used as a complement to formal verification to verify implementation details that are ignored by the models.

\section{Conclusion}

In this paper, we presented \name{}, a novel pre-silicon processor fuzzer designed to detect transient execution vulnerabilities effectively and efficiently.
\name{} introduces two innovative operating primitives to enhance microarchitectural controllability and observability.
By leveraging dynamic swappable memory and differential information flow tracking, \name{} efficiently triggers diverse transient windows, effectively guides mutation, and identifies exploitable leakages.
We evaluated \name{} on two well-known RISC-V out-of-order processors and achieved up to 4.7$\times$ improvement in coverage compared to the state-of-the-art fuzzer \specdoc{}.
Moreover, \name{} identified 5 new transient execution vulnerabilities (with 6 CVEs assigned), showing its effectiveness in detecting previously unknown bugs.



\end{document}